\documentclass{iaus}
\usepackage{graphicx}
\newcommand{\msun}{\mbox{M$_{\odot}$}}
\newcommand{\rsun}{\mbox{R$_{\odot}$}}

\title[Distances with Eclipsing Binaries] 
{Eclipsing Binaries: Tools for Calibrating the Extragalactic Distance Scale}

\author[A.Z. Bonanos]   %% give here short author list %%
{Alceste Z. Bonanos}

\affiliation{Carnegie Institution of Washington, 5241 Broad Branch Road,
  Washington, DC 20015, USA \break email: bonanos@dtm.ciw.edu}

\pubyear{2006}
\volume{240}  %% insert here IAU Symposium No.
\pagerange{119--126}
\date{?? and in revised form ??}
\jname{Binary Stars As Critical Tools \& Tests}
\editors{W. Hartkopf, E. Guinan \& P. Harmanec, eds.}
\begin{document}

\maketitle

\begin{abstract}
In the last decade, over 7000 eclipsing binaries have been discovered in
the Local Group through various variability surveys. Measuring
fundamental parameters of these eclipsing binaries has become feasible
with 8 meter class telescopes, making it possible to use eclipsing
binaries as distance indicators. Distances with eclipsing binaries
provide an independent method for calibrating the extragalactic distance
scale and thus determining the Hubble constant. This method has been
used for determining distances to eclipsing binaries in the Magellanic
Clouds and the Andromeda Galaxy and most recently to a detached
eclipsing binary in the Triangulum Galaxy by the DIRECT Project. The
increasing number of eclipsing binaries found by microlensing and
variability surveys also provide a rich database for advancing our
understanding of star formation and evolution.

\end{abstract}

\firstsection % if your document starts with a section,
              % remove some space above using this command.
\section{Introduction}

The last decade has seen a dramatic increase in the number of
extragalactic eclipsing binaries discovered. Most of these have been
found as a side product of the microlensing surveys towards the Large
and Small Magellanic Clouds (LMC and SMC). Starting in the early 1990s
the EROS Experiment, the MACHO Project, the Microlensing Observations in
Astrophysics (MOA) and the OGLE Project began monitoring the Magellanic
Clouds with 1 meter telescopes in a search for dark matter in the form
of massive compact halo objects. As a side product they have discovered
thousands of variable stars, including many eclipsing binaries. The 75
EROS binaries in the LMC published by \cite{Grison95} doubled the known
binaries in the galaxy at the time. The Microlensing Observations in
Astrophysics (MOA) group soon after released a catalog of 167 eclipsing
binaries in the SMC (\cite[Bayne et al. 2002]{Bayne02}), followed by the
extensive catalogs of the OGLE-II Project which comprise of 2580
eclipsing binaries in the LMC (\cite[Wyrzykowski et
al. 2004]{Wyrzykowski03}) and 1350 in the SMC (\cite[Wyrzykowski et
al. 2004]{Wyrzykowski04}). A catalog of the eclipsing binaries found by
the MACHO project with $\sim4500$ binaries in the LMC and 1500 in the
SMC is underway. Note, that some of these binaries are foreground or
galactic. The SuperMACHO project, using the CTIO Blanco 4 meter
telescope, has surveyed the Magellanic Clouds down to $VR\sim23$ mag
(\cite[Huber et al. 2005]{Huber2005}), extending the sample to include
solar type stars, some of which will be W UMa variables. The first
extragalactic W UMa detection was recently made by \cite{Kaluzny06},
with photometry from the 6.5 meter Magellan telescopes at Las Campanas,
Chile.

Fewer eclipsing binaries are known in more distant galaxies, partly due
to their faintness and the necessity of large amounts of time on medium
size telescopes (2-4 meters) to obtain good quality light curves. Two
microlensing surveys are underway toward M31. The Wendelstein Calar Alto
Pixellensing Project (WeCAPP, \cite[Fliri et al. 2006]{Fliri06})
discovered 31 eclipsing binaries in the bulge of M31. The variable star
catalog released by the POINT-AGAPE Survey (\cite[An et al. 2004]{An04})
has not been searched systematically for eclipsing binaries, but is
bound to contain some among the 35000 variables.

The DIRECT Project (see Stanek et al. 1998; Bonanos et al. 2003) began
monitoring M31 and M33 in 1996, specifically for Cepheids and detached
eclipsing binaries (DEBs) with the 1.2 meter telescope on Mt Hopkins,
Arizona. In M31, a total of 89 eclipsing binaries were found in the 6
fields surveyed. A variability survey using the 2.5 meter Isaac Newton
telescope by \cite{Vilardell06} has found 437 eclipsing binaries in M31,
bringing the total to over 550 eclipsing binaries. In M33, the DIRECT
Project has found 148 eclipsing binaries (see \cite[Mochejska et
al. 2001]{Mochejska01}, and references within).  Eclipsing binaries have
also been discovered in NGC 6822, a dwarf irregular galaxy in the Local
Group, by the Araucaria Project (\cite[Mennickent et
al. 2006]{Mennickent06}).

It is worth mentioning the eclipsing binaries discovered beyond the
Local Group. The first such discovery was made back in 1968 by
\cite[Tammann \& Sandage]{Tammann68}, who presented the light curve of a
6 day period binary in NGC 2403 (M81 group) with $B\sim22$ mag. More
recently, the Araucaria Project has discovered a binary in NGC 300
(Sculptor group). \cite{Mennickent04} present the light curve of the
$B\sim21.5$ mag detached eclipsing binary in this galaxy.

Finally, the discovery of 3 Cepheid binaries in the LMC by
\cite{Udalski99, Alcock02} provides a new way of calibrating the Cepheid
period-luminosity relation and the extragalactic distance scale.

\section{Eclipsing Binaries as Distance Indicators}

Eclipsing binaries provide an accurate method of measuring distances to
nearby galaxies with an unprecedented accuracy of 5\% -- a major step
towards a very accurate and independent determination of the Hubble
constant. Reviews and history of the method can be found in
\cite[Andersen (1991)]{Andersen91} and \cite{Paczynski97}. The method
requires both photometry and spectroscopy of an eclipsing binary. From
the light and radial velocity curve the fundamental parameters of the
stars can be determined accurately. The light curve provides the
fractional radii of the stars, which are then combined with the
spectroscopy to yield the physical radii and effective temperatures. The
velocity semi-amplitudes determine both the mass ratio and the sum of
the masses, thus the individual masses can be solved for. Furthermore,
by fitting synthetic spectra to the observed ones, one can infer the
effective temperature, surface gravity and luminosity. Comparison of the
luminosity of the stars and their observed brightness yields the
reddening of the system and distance.

Measuring distances with eclipsing binaries is an essentially geometric
method and thus accurate and independent of any intermediate calibration
steps. With the advent of 8 m class telescopes, eclipsing binaries have
been used to obtain accurate distance estimates to the LMC, SMC, M31 and
M33; these results are presented below.

\section{Eclipsing Binary Distances to the Magellanic Clouds and M31}

The first extragalactic distance measurement using a detached eclipsing
binary system was published by \cite{Guinan98}, demonstrating the
importance of EBs as distance indicators. The detached system 14$^{th}$
mag system HV 2274 was observed with the Faint Object Spectrograph (FOS)
onboard the \textit{Hubble Space Telescope}. The UV/optical
spectrophotometry was used to derive the radial velocity curve and
reddening. The distance to HV 2274 was determined to be $47.0\pm2.2$ kpc
(\cite[Guinan et al. 1998, Fitzpatrick et al. 2002]{Guinan98,
Fitzpatrick02}). Distances to 3 more systems in the LMC have been
determined: the detached 15$^{th}$ magnitude system HV 982
(\cite{Fitzpatrick02}) at a distance of $50.2\pm1.2$ kpc, the 15$^{th}$
magnitude detached system EROS 1044 (\cite{Ribas02}) at $47.5\pm1.8$ kpc
and the 14$^{th}$ magnitude semi-detached system HV 5936 at $43.2\pm1.8$
kpc (\cite{Fitzpatrick03}). These support a ``short'' distance scale to
the LMC, in contrast to the LMC distance of 50 kpc adopted by the Key
Project (\cite{Freedman01}.  The spread in the distances is most likely
an indication of the intrinsic extent of the LMC along the line of
sight.

\cite{Harries03} and \cite{Hilditch05} have conducted a systematic
spectroscopic survey of eclipsing binaries in the SMC, obtaining
fundamental parameters and distances to 50 eclipsing binary
systems. Their sample was selected from the OGLE-II database of SMC
eclipsing binaries as the brightest systems ($B<16$ mag) with short
periods ($P_{orb}<5$ days) to increase the efficiency of multi-fiber
spectroscopy over a typical observing run. The mean true distance
modulus from the whole sample is
$18.91\pm0.03(random)\pm0.1(systematic)$ and the implied LMC distance is
$18.41\pm0.04(random)\pm0.1(systematic)$, again in support of the
``short'' distance scale.

M31 and M33, being the nearest spiral galaxies, are crucial
stepping-stones in the extragalactic distance ladder. \cite{Ribas05}
have determined the first distance to a spiral galaxy, specifically to a
semi-detached system ($V=19.3$ mag) in M31. Note that at such distances,
the location of the binary within the galaxy has an insignificant effect
on the distance ($<1\%$). Light curves for the system in M31 were
obtained from the survey of \cite{Vilardell06} with the 2.5 meter Isaac
Newton telescope and spectroscopy with the 8 meter Gemini telescope
using GMOS. Such stars are at the limit of current spectroscopic
capabilities. The resulting distance is $772\pm44$ kpc and distance
modulus is $24.44\pm0.12$ mag, in agreement with previous distance
determinations to M31.

\section{DIRECT Distance to a Detached Eclipsing Binary in M33}

\subsection{Motivation}

The DIRECT Project (see Stanek et al. 1998; Bonanos et al. 2003) aims to
measure distances to the nearby Andromeda (M31) and Triangulum (M33)
galaxies with eclipsing binaries and the Baade-Wesselink method for
Cepheids. It began surveying these galaxies in 1996 with 1 m class
telescopes. The goal of the DIRECT Project is to replace the current
anchor galaxy of the extragalactic distance scale, the LMC, with the
more suitable spiral galaxies in the Local Group, M31 and M33. These are
the nearest spiral galaxies to ours, yet more than ten times more
distant than the LMC and therefore more difficult to observe stars in
them. The Cepheid period-luminosity relation is used to measure
distances to a few tens of Mpc, while Type Ia supernovae are used to
probe distances out to a few hundred Mpc. Galaxies hosting both Cepheids
and Type Ia supernovae become calibrators of the luminosities of
supernovae, which are used to determine the Hubble constant, $H_0$.

How is the Cepheid period-luminosity calibrated? Benedict et al. (2002)
in their Figure 8 show 84 recent measurements of the distance modulus of
the LMC using 21 methods. The large spread in the different measurements
is quite disturbing. There are several problems with using the LMC as
the anchor of the distance scale, which demand its replacement. The zero
point of the period-luminosity relation is not well determined and the
dependence on metallicity remains controversial. There is increasing
evidence for elongation of the LMC along the line of sight that
complicates a distance measurement. One has to additionally include a
model of the LMC when measuring distances, which introduces systematic
errors. Finally, the reddening across the LMC has been shown to be
variable (Nikolaev et al. 2004), which has to be carefully accounted
for. These effects add up to a 10-15\% error in the distance to the LMC,
which in the era of precision cosmology is unacceptable. The replacement
of the current anchor galaxy of the distance scale with a more suitable
galaxy or galaxies is long overdue. Furthermore, the {\it Hubble Space
Telescope} Key Project (Freedman et al. 2001) has measured the value of
$H_0$ by calibrating Cepheids measured in spiral galaxies and secondary
distance indicators and found $H_{0}=72\pm8\; \rm km\;s^{-1}\;
Mpc^{-1}$. This result is heavily dependent on the distance modulus to
the LMC they adopt (18.50 mag or 50 kpc).

\subsection{DIRECT Observations}

The DIRECT project involves three stages: surveying M31 and M33 in order
to find detached eclipsing binaries and Cepheids; once discovered
selecting and following up the best targets with medium size telescopes
(2-4 m class) to obtain more accurate light curves and lastly, obtaining
spectroscopy which requires 8-10 m class telescopes. DIRECT completed
the survey stage in 1996-1999 with 200 full/partial nights on 1 m class
telescopes in Arizona. Follow up observations of the 2 best eclipsing
binaries were obtained in 1999 and 2001 using the Kitt Peak 2.1 m
telescope in Arizona. The total number of eclipsing binaries found in
M33 were 237, however only 4 are bright enough ($V_{max}<20$ mag) for
distance determination with currently available telescopes. The criteria
for selection include a detached configuration (stars are well within
their Roche lobes) and deep eclipses, which remove degeneracies in the
modeling and a short period ($<10$ days) that makes follow up
observations feasible.

Bonanos et al. (2006) presented the first distance determination to a
detached eclipsing binary (DEB) in M33 that was found by Macri et
al. (2001). D33J013346.2+304439.9 is located in the OB 66
association. Follow up optical data were obtained in order to improve
the quality of the light curve and additional infrared observations were
made using the 8 m Gemini telescope in order to better constrain the
extinction to the system. Spectra of the DEB were obtained in 2002-2004
with the 10-meter Keck-II telescope and 8 m Gemini telescope on Mauna
Kea. Note that $\sim4$ hours of observations per epoch were required for
radial velocity measurements, a large investment of 8-10 m class
telescope time. Absorption lines from both stars are clearly resolved in
the spectrum, making it a double lined spectroscopic binary.

Careful modeling with non-local thermodynamic equilibrium model spectra
yielded effective temperatures T$_{\rm eff1}=37000\pm1500$~K and T$_{\rm
eff2}=35600\pm1500$~K. The primary star is defined as the hotter star
eclipsed at phase zero. We measured radial velocities from the spectra
and from the light and radial velocity curves derived the parameters of
the DEB components. The $V-$band light curve model fit for the DEB is
shown in Figure~1. Note that the deviation of the secondary eclipse from
phase 0.5 is due to the eccentricity of the system. The radial velocity
curve is presented in Figure~2. The rms residuals are 26.0 $\rm
km\;s^{-1}$ for the primary and 28.0 $\rm km\;s^{-1}$ for the secondary
star. We find the DEB components to be O7 type stars with masses: $\rm
M_{1}=33.4\pm3.5 \; \msun$, $\rm M_{2}=30.0\pm3.3\; \msun$ and radii
$\rm R_{1}=12.3\pm0.4\; \rsun$, $\rm R_{2}=8.8\pm0.3\; \rsun$.

\subsection{Distance Determination}

Having measured the temperatures of the stars from the spectra, we
computed fluxes and fit the optical and near-infrared $BVRJHKs$
photometry. The best fit that minimized the photometric error over the 6
photometric bands yielded a distance modulus to the DEB and thus M33 of
$24.92\pm0.12$ mag ($964\pm54$ kpc). The fit of the reddened model
spectrum to the photometry is shown in Figure~3.

There are several avenues for improving the distance to M33 and M31
using eclipsing binaries. \cite{Wyithe02} propose the use of
semi-detached eclipsing binaries to be just as good or better distance
indicators as detached eclipsing binaries, which have been traditionally
considered to be ideal. Semi-detached binaries provide other benefits:
their orbits are tidally circularized and their Roche lobe filling
configurations provide an extra constraint in the parameter space,
especially for complete eclipses. Bright semi-detached binaries in M33
or M31 are not as rare as DEBs, and are easier to follow-up
spectroscopically, as demonstrated by \cite{Ribas05} in M31. Thus, for
the determination of the distances to M33 and M31 to better than $5\%$
we suggest both determining distances to other bright DEBs and to
semi-detached systems found by DIRECT and other variability
surveys. Additional spectroscopy of the DEB would also improve the
current distance determination to M33, since the errors are dominated by
the uncertainty in the radius or velocity semi-amplitude.

How does our M33 distance compare to previous determinations?  Table~1
(adapted from Bonanos et al. 2006) presents a compilation of 13 recent
distance determinations to M33 ranging from 24.32 to 24.92 mag,
including the reddening values used. Our measurement although completely
independent yields the largest distance with a small 6\% error, thus is
not consistent with some of the previous determinations. This possibly
indicates unaccounted sources of systematic error in the calibration of
certain distance indicators. Note the Freedman et al. (2001) distance to
M33 is not consistent with the DIRECT measurement. This could be due to
their ground based photometry which is likely affected by blending, but
highlights the importance of securing the anchor of the extragalactic
distance scale. The eclipsing binary distances to the LMC presented
above indicate a shorter distance to the LMC. Combined with eclipsing
binary distances to M31 and M33, we should soon be able to reduce the
errors in the distance scale and thus the Hubble constant to 5\% or
better.

\begin{table}[h]
\caption{\sc Recent Distance Determinations to M33}
 \begin{tabular}{lccc}\hline
Study & Method$^{\rm a}$ & Distance Modulus & Reddening \\\hline%\\%[3pt]
\cite{Bonanos06} & DEB & $24.92\pm0.12$ & $E(B-V)=0.09\pm0.01$ \\
\cite{Sarajedini06} & RR Lyrae & $24.67\pm0.08$ & $\sigma_{E(V-I)}=0.30$\\
\cite{Brunthaler05} & Water Masers & $24.32\pm0.45$ & --- \\
\cite{Ciardullo04} & PNe & $24.86^{+0.07}_{-0.11}$ & $E(B-V)=0.04$ \\
\cite{Galleti04} & TRGB & $24.64\pm0.15$ &$E(B-V)=0.04$  \\
\cite{McConnachie04} & TRGB & $24.50\pm0.06$ &$E(B-V)=0.042$ \\
\cite{Tiede04} & TRGB & $24.69\pm0.07$ &$E(B-V)=0.06\pm0.02$ \\
\cite{Kim02} & TRGB & $24.81\pm0.04(r)^{+0.15}_{-0.11}(s)$ & $E(B-V)=0.04$\\
\cite{Kim02} & RC & $24.80\pm0.04(r)\pm0.05(s)$ & $E(B-V)=0.04$ \\
\cite{Lee02} & Cepheids & $24.52\pm0.14(r)\pm0.13(s)$ &
$E(B-V)=0.20\pm0.04$ \\
\cite{Freedman01} & Cepheids & $24.62\pm0.15$ & $E(V-I)=0.27$ \\
\cite{Pierce00} & LPVs & $24.85\pm0.13$ & $E(B-V)=0.10$ \\
\cite{Sarajedini00} & HB & $24.84\pm0.16$ & $<E(V-I)>=0.06\pm0.02$ \\\hline
\end{tabular}
\label{distances}
\footnote{}{DEB: detached eclipsing binary; TRGB: tip of the red
giant branch; PNe: planetary nebulae; RC: the red clump; LPVs: long
period variables; HB: horizontal branch stars.}
\end{table}

\subsection{Epilogue}

The accelerating rate of discovery of eclipsing binaries provides
immense opportunities. With current spectroscopic capabilities it has
become possible to measure distances to Local Group galaxies out to 1
Mpc, thus providing distances independent of the controversial LMC
distance and the calibration of Cepheids, which most methods rely on. An
independent calibration of the extragalactic distance scale has become
possible. The recent distance determinations to the LMC, SMC, M31 and
M33 are providing 6\% distances to these galaxies that will improve over
the next few years. Combined with geometric distances to the maser
galaxy NGC 4258, the extragalactic distance scale will soon be anchored
to several spiral galaxies (M31, M33, NGC 4258).

In addition to their use as distance indicators, eclipsing binaries
provide many more opportunities to advance our understanding of star
formation and evolution. In particular, they provide direct means of
measuring masses, radii and luminosities of stars. For example,
applications to the extremes of stellar mass ranges are underway in
order to provide constraints to theoretical models of stellar
atmospheres and evolution, such as for M-dwarfs and at the other extreme
for very massive ($>50\;\msun$) O-stars and Wolf-Rayet stars. Future
projects such as the wide field imaging surveys Pan-STARRS and the Large
Synoptic Survey Telescope (LSST) will survey the sky down to 24th mag
and yield thousands of binaries in the Galaxy, the Local Group and
beyond.

\begin{figure}
\includegraphics[width=7in]{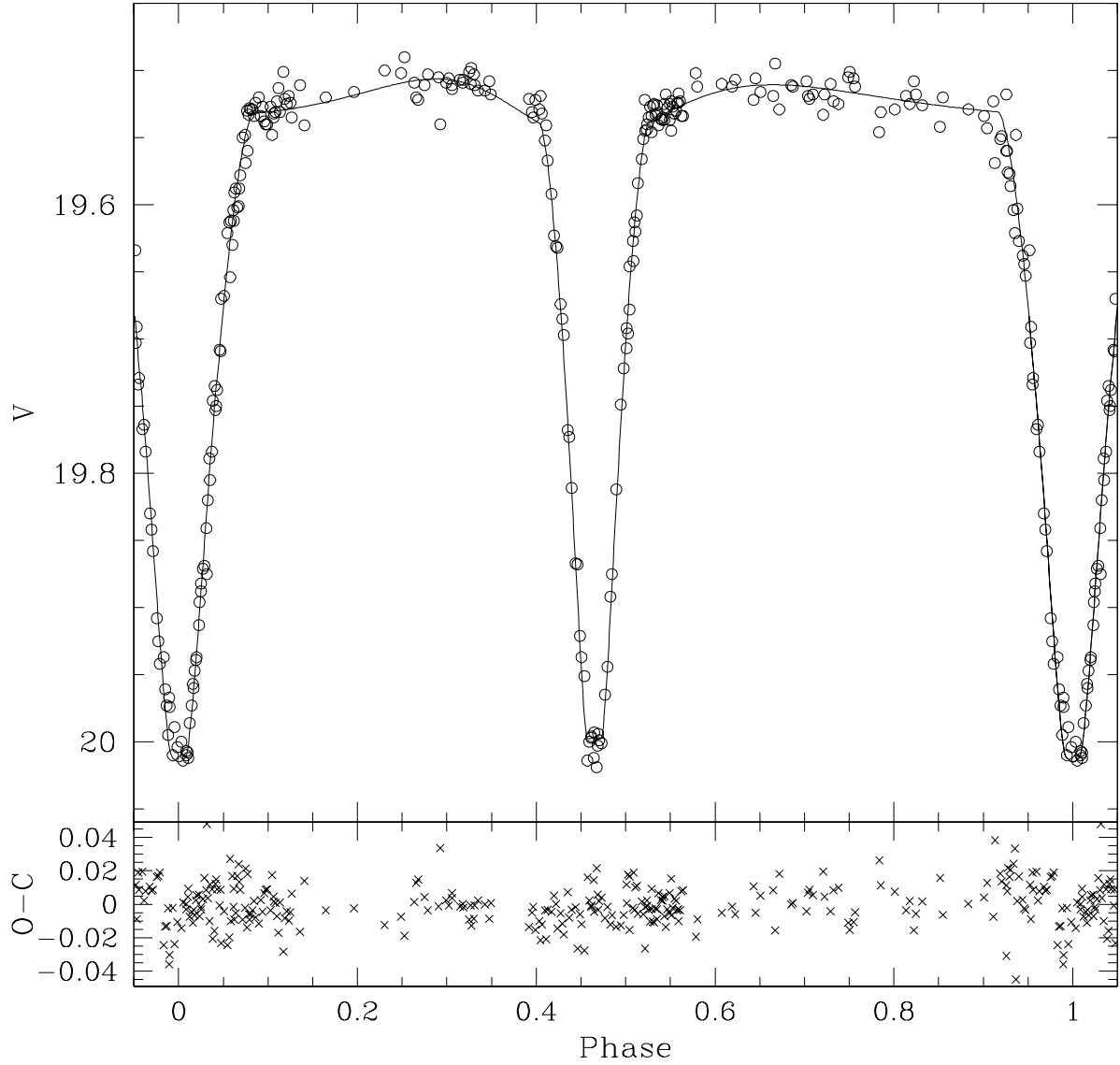}
\caption{$V-$band light curve of the M33 DEB with model fit from the
Wilson-Devinney program. Circles correspond to the 278 V-band
observations and the solid line to the model; the rms is 0.01 mag (from
Bonanos et al. 2006).}
\end{figure}

\begin{figure}
\includegraphics[width=7in]{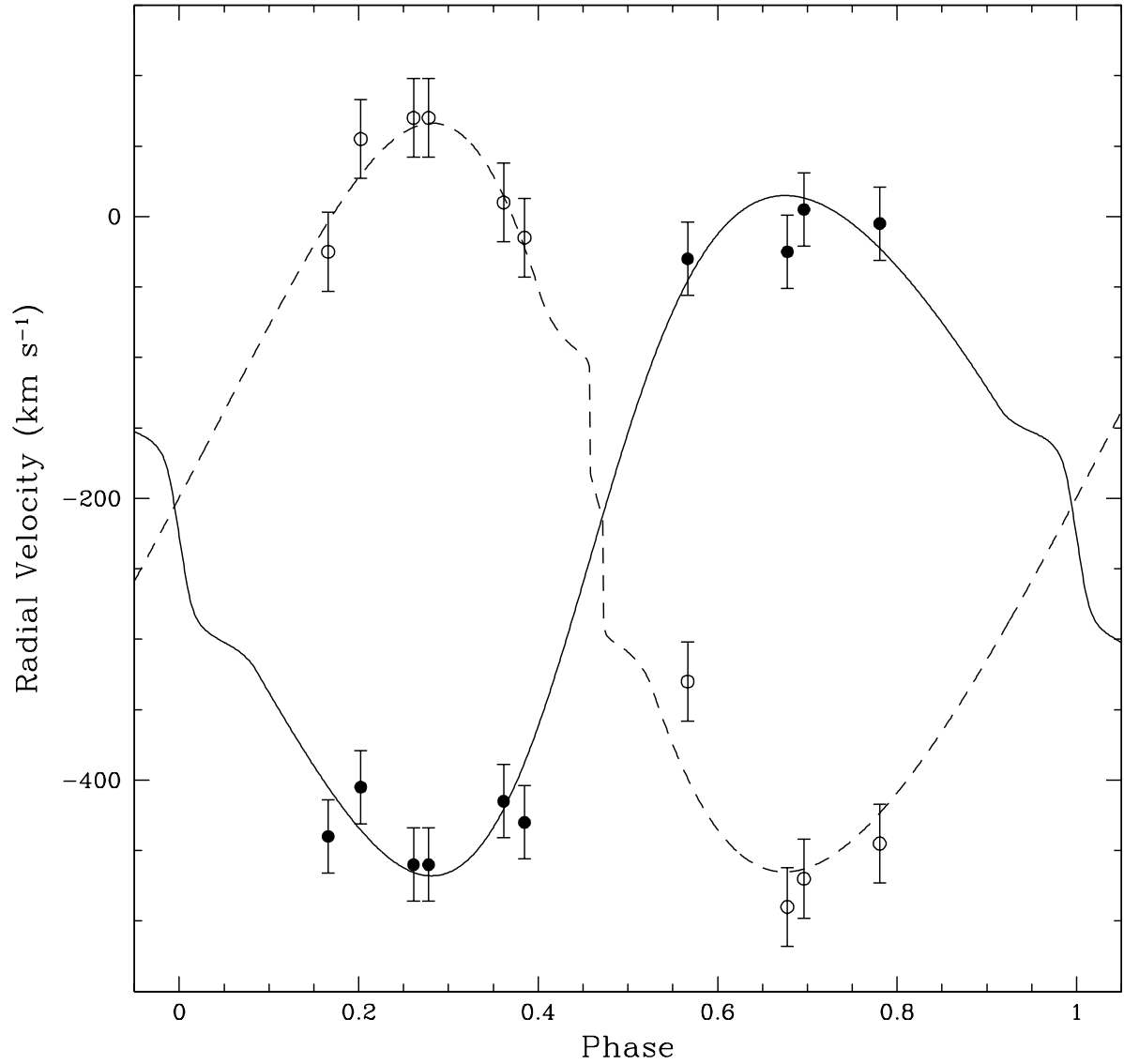}
\caption{Radial velocities for the DEB measured by two-dimensional cross
correlation with synthetic spectra. Model fit is from Wilson-Devinney
program. Error bars correspond to the rms of the fit: 26.0 $\rm km\;
s^{-1}$ for the primary (filled circles) and 28.0 $\rm km\; s^{-1}$ for
the secondary (open circles).}
\end{figure}

\clearpage
\begin{figure}
\includegraphics[angle=90, width=7in]{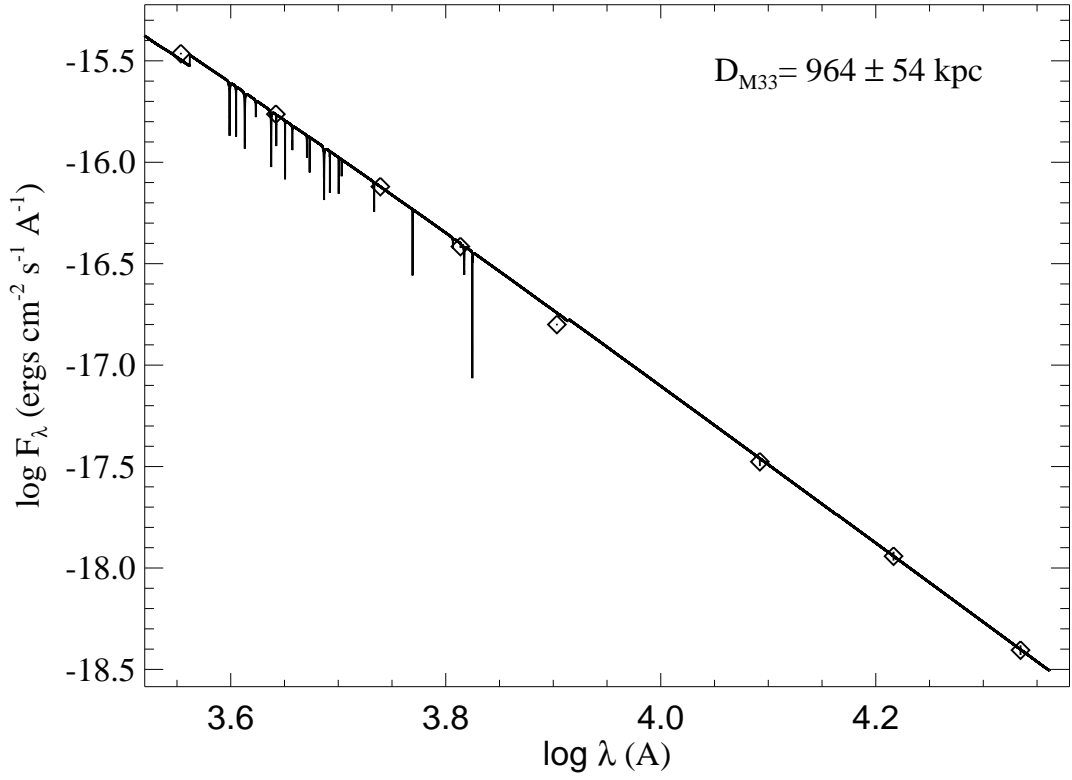}
\caption{Fit of the reddened DEB model spectrum to the $BVRJHK_s$
ground-based photometry. Overplotted is the $U$ and $I$ photometry from
\cite{Massey06}. The distance modulus to the DEB and thus M33 is found
to be $24.92\pm0.12$ mag ($964\pm54$ kpc).}
\end{figure}


\begin{thebibliography}{}

\bibitem[Alcock \etal\ (2002)]{Alcock02}
     {Alcock, C., Allsman, R.A., Alves, D.R., et al.} 2002,
     \textit{ApJ}, 573, 338

\bibitem[An et al. (2004)]{An04}
     {An, J.H., Evans, N.W., Hewett, P., et al.} 2004,
     \textit{MNRAS} 351, 1071

\bibitem[Bayne et al. (2002)]{Bayne02}
     {Bayne, G., Tobin, W., Bond, I., et al.} 2002,
     \textit{MNRAS}, 331, 609

\bibitem[Benedict et al. (2002)]{Benedict02}
Benedict, G.F., McArthur, B.E., Fredrick, L.W., {et~al.} 2002, \textit{AJ}, 123, 473

\bibitem[Bonanos et al. (2003)]{Bonanos03}
     {Bonanos, A.Z., Stanek, K.Z., Sasselov, D.D., et al.} 2003,
     \textit{AJ}, 126, 175

\bibitem[Bonanos et al. (2006)]{Bonanos06}
     {Bonanos, A.Z., Stanek, K.Z., Kudritzki, R.P., et al.} 2006,
     \textit{ApJ}, 652, 313

\bibitem[{{Brunthaler} \etal\ (2005)}]{Brunthaler05}
{Brunthaler}, A., {Reid}, M.~J., {Falcke}, H., {et~al.} 2005, \textit{Science}, 307,
  1440

\bibitem[{{Ciardullo} {et~al.} (2004)}]{Ciardullo04}
{Ciardullo}, R., {Durrell}, P.~R., {Laychak}, M.~B., {et~al.} 2004, \textit{ApJ}, 614,
  167

\bibitem[Fitzpatrick {et~al.} 2002]{Fitzpatrick02}
{Fitzpatrick}, E.~L., {Ribas}, I., {Guinan}, E.~F., {et~al.} 2002,
\textit{ApJ}, 564, 260

\bibitem[Fitzpatrick {et~al.} 2003]{Fitzpatrick03}
{Fitzpatrick}, E.~L., {Ribas}, I., {Guinan}, E.~F., {et~al.} 2003, \textit{ApJ}, 587,
  685

\bibitem[Fliri \etal\ (2006)]{Fliri06}
     {Fliri, J., Riffeser, A., Seitz, S., \& Bender, R.} 2006,
     \textit{A\&A}, 445, 423

\bibitem[Freedman {et~al.} 2001]{Freedman01}
{Freedman}, W.~L., {Madore}, B.~F., {Gibson}, B.~K., {et~al.} 2001, \textit{ApJ}, 553,
  47

\bibitem[{{Galleti} {et~al.} (2004)}]{Galleti04}
{Galleti}, S., {Bellazzini}, M., \& {Ferraro}, F.~R. 2004, \textit{A\&A}, 423, 925

\bibitem[Grison \etal\ (1995)]{Grison95}
     {Grison, P., Beaulieu, J.-P., Pritchard, J.D., et al.} 1995,
     \textit{A\&AS}, 109, 447

\bibitem[{{Guinan} {et~al.} (1998)}]{Guinan98}
{Guinan}, E.~F., {Fitzpatrick}, E.~L., {Dewarf}, L.~E., {et~al.} 1998, \textit{ApJ}, 509, L21

\bibitem[Harries \etal\ (2003)]{Harries03}
     {Harries, T.J., Hilditch, R.W., \& Howarth, I.D.} 2003,
     \textit{MNRAS}, 339, 157

\bibitem[Hilditch \etal\ (2005)]{Hilditch05}
     {Hilditch, R.W., Howarth, I.D., \& Harries, T.J.} 2005,
     \textit{MNRAS}, 357, 304

\bibitem[Huber \etal\ (2005)]{Huber05}
     {Huber, M.E., Nikolaev, S., Cook, K.H., et al.} 2005,
     \textit{BAAS}, 207, 122.16

\bibitem[Kaluzny \etal\ (2006)]{Kaluzny06}
     {Kaluzny, J., Mochnacki, S., \& Rucinski, S.M.} 2006,
      \textit{AJ}, 131, 407

\bibitem[{{Kim} {et~al.} (2002)}]{Kim02}
{Kim}, M., {Kim}, E., {Lee}, M.~G., {et~al.} 2002, \textit{AJ}, 123, 244

\bibitem[{{Lee} {et~al.} (2002)}]{Lee02}
{Lee}, M.~G., {Kim}, M., {Sarajedini}, A., {et~al.} 2002, \textit{ApJ}, 565, 959

\bibitem[Macri \etal\ 2001]{Macri01}
     {Macri, L.M., Stanek, K.Z., Sasselov, D.D., et al.} 2001,
     \textit{AJ}, 121, 870 

\bibitem[Massey \etal\ (2006)]{Massey06}
     {Massey, P., Olsen, K.A.G., Hodge, P.W. et al.} 2006,
     \textit{AJ}, 131, 2478

\bibitem[{{McConnachie} {et~al.} (2004)}]{McConnachie04}
{McConnachie}, A.~W., {Irwin}, M.~J., {Ferguson}, A.~M.~N., {et~al.} 2004,
  \textit{MNRAS}, 350, 243

\bibitem[Mennickent \etal\ (2004)]{Mennickent04}
     {Mennickent, R.E., Pietrzynski, G., \& Gieren, W.} 2004,
     \textit{MNRAS}, 350, 679

\bibitem[Mennickent \etal\ (2006)]{Mennickent06}
     {Mennickent, R.E., Gieren, W., Soszynski, I., \& Pietrzynski, G.} 2006,
     \textit{A\&A}, 450, 873

\bibitem[Mochejska \etal\ (2001)]{Mochejska01}
     {Mochejska, B.J., Kaluzny, J., Stanek, K.Z., et al.} 2001,
      \textit{AJ}, 122, 2477

\bibitem[Nikolaev \etal\ (2004)]{Nikolaev04}
Nikolaev, S., Drake, A.J., Keller, S.C., {et~al}. 2004, \textit{ApJ}, 601, 260

\bibitem[Paczynski (1997)]{Paczynski97}
     {Paczynski, B.} 1997
      \textit{The Extragalactic Distance Scale} 273

\bibitem[{{Pierce} {et~al.} (2000)}]{Pierce00}
{Pierce}, M.~J., {Jurcevic}, J.~S., \& {Crabtree}, D. 2000, \textit{MNRAS}, 313, 271

\bibitem[{{Ribas} {et~al.} 2002}]{Ribas02}
{Ribas}, I., Fitzpatrick, E.L., Malonley, F.P., {et~al.} 2002, 
\textit{ApJ}, 574, 771

\bibitem[{{Ribas} {et~al.} (2005)}]{Ribas05}
{Ribas}, I., {Jordi}, C., {Vilardell}, F., {et~al.} 2005, \textit{ApJ}, 635, L37

\bibitem[{{Sarajedini} {et~al.} (2000)}]{Sarajedini00}
{Sarajedini}, A., {Geisler}, D., {Schommer}, R., {et~al.} 2000, \textit{AJ}, 120, 2437

\bibitem[{Sarajedini} {et~al.} (2006)]{Sarajedini06}
{Sarajedini}, A., {Barker}, M., {Geisler}, D., {et~al.} 2006, \textit{AJ}, 132, 1361

\bibitem[{Stanek} {et~al.} (1998)]{Stanek98}
Stanek, K.Z., Kaluzny, J., Krockenberger, M., {et~al.} 1998, \textit{AJ}, 115, 1894 

\bibitem[Tammann \etal\ (1968)]{Tammann68}
     {Tammann, G.A., \& Sandage, A.} 1968,
     \textit{ApJ}, 151, 825

\bibitem[{{Tiede} {et~al.} (2004)}]{Tiede04}
{Tiede}, G.~P., {Sarajedini}, A., \& {Barker}, M.~K. 2004, \textit{AJ}, 128, 224

\bibitem[Udalski \etal\ (1999)]{Udalski99}
     {Udalski, A., Soszynski, I., Szymanski, M., et al.} 1999,
     \textit{AcA}, 49, 223 

\bibitem[Vilardell \etal\ (2006)]{Vilardell06}
     {Vilardell, F., Ribas, I., \& Jordi, C.} 2006,
     \textit{A\&A}, in press (astro-ph/0607236)

\bibitem[{{Wyithe} \& {Wilson} (2002)}]{Wyithe02}
{Wyithe}, J.~S.~B., \& {Wilson}, R.~E. 2002, \textit{ApJ}, 571, 293

\bibitem[Wyrzykowski \etal\ (2003)]{Wyrzykowski03}
     {Wyrzykowski, L., Udalski, A., Kubiak, M., et al.} 2003,
     \textit{AcA}, 53, 1

\bibitem[Wyrzykowski \etal\ (2004)]{Wyrzykowski04}
     {Wyrzykowski, L., Udalski, A., Kubiak, M., et al.} 2004,
     \textit{AcA}, 54, 1

\end{thebibliography}
\end{document}